\documentclass[fleqn,10pt]{okumura} \title{Meandering instability of air flow in a granular bed: self-similarity and fluid-solid duality} \author[]{Yuki Yoshimura} \author[]{Yui Yagisawa} \author[1,*]{Ko Okumura} \affil[]{Department of Physics and Soft Matter Center, Ochanomizu University, 2--1--1, Otsuka, Bunkyo-ku, Tokyo 112-8610, Japan} \affil[*]{Corresponding author: okumura@phys.ocha.ac.jp} 

\begin{abstract}
Meandering instability is familiar to everyone through river meandering or
small rivulets of rain flowing down a windshield. However, its physical
understanding is still premature, although it could inspire researchers in
various fields, such as nonlinear science, fluid mechanics and geophysics, to
resolve their long-standing problems. Here, we perform a small-scale
experiment in which air flow is created in a thin granular bed to successfully
find a meandering regime, together with other remarkable fluidized regimes,
such as a turbulent regime. We discover that phase diagrams of the flow
regimes for different types of grains can be universally presented as
functions of the flow rate and the granular-bed thickness when the two
quantities are properly renormalized. We further reveal that the meandering
shapes are self-similar as was shown for meandering rivers. The experimental
findings are explained by theory, with elucidating the physics. The theory is
based on force balance, a minimum-dissipation principle, and a
linear-instability analysis of a continuum equation that takes into account
the fluid-solid duality, i.e., the existence of fluidized and solidified
regions of grains along the meandering path. The present results provide
fruitful links to related issues in various fields, including fluidized bed
reactors in industry.

\end{abstract}
\begin{document}

\flushbottom\maketitle
\thispagestyle{empty}

{Fluid }flows in a medium often display spectacular instabilities; there are
various examples of this in cosmological \cite{CosmicHelixScience2001},
geophysical \cite{MantleTransportNature1988}, biological
\cite{kessler1986JFM86}, and physical systems
\cite{CoilLiquidPRLGoldstein2005}. Among such phenomena, meandering
instability is familiar to everyone in the form of small rivulets of rain
flowing down a windshield or, in a more large-scale form, in river meandering.
In particular, for a meandering fluid interacting with a solid surface, the
physics has been relatively well understood. The meandering of a liquid jet
flowing down an inclined plate and some variations of this have actively been
studied\cite{Nakagawa1984,braiding,Drenckhan2004} (for larger-scale
experiments, see Ref.\cite{Braudrick2009} and references therein). For such a
meandering, the physical mechanism has been elucidated in terms of force
balance, linear stability \cite{LimatPRL06,LimatPRL2011,DaerrEPL2012} and
flow-rate fluctuation\cite{Birnir2008}. On the contrary, for a meandering
fluid interacting with a surrounding "complex fluid," the mechanism of the
instability has yet to be clarified. This latter case includes the meandering
of rivers, which is ubiquitous and is even found on Mars \cite{Howard2009},
and has been studied in geography \cite{Braudrick2009,Lewin1972}, biology
\cite{Trinkaus1998,Napieralski1996,Flierl1993}, hydrodynamics
\cite{Sawai1981,Howard1984,Stolum1996}, and physics, from various viewpoints
such as pattern formation \cite{Stolum1996,SwinneyPRL96}, random walks
\cite{Howard1970,Ferguson1976}, and statistical models
\cite{Bruinsma1990,EdwardsMeander}. For example, more than a half-century ago,
shapes of meandering rivers were shown to be scale invariant or self-similar,
i.e., the curvature, amplitude, and wavelength of the shape all scale with the
width, using the data obtained by field studies \cite{MeanderLeopold}, which
has yet to be understood. More recently, a systematic dependence of the
sinuosity of meandering rivers on the Froude number has been shown, which is
interpreted through results obtained by a simple numerical model
\cite{SinuosityPNAS2013}.

Here, we perform a small-scale experiment in which an air flow is created in a
granular bed as follows: the flow of a lighter fluid is surrounded by a
heavier "complex fluid," which is similar to river meandering. As a result, we
find that the flow can be destabilized to show meandering shapes. The
meandering regime can be universally demonstrated as a function of the
normalized flow rate and normalized granular-bed thickness. In addition, the
meandering flow shape is found to be scale invariant or self-similar, as was
shown for river meandering. These experimental findings can be explained by
the proposed physical principles.

\section*{Results}

\subsection*{Experiment}

\subsubsection*{Setup}

The main setup consists of a transparent acrylic cell and a gas-flow
controller (see Fig.~\ref{Fig1}(a)). The inside thickness $D$, width $W$ and
height $H$ of the cell satisfy the condition $d\ll D\ll W<H$. The cell was
filled with beads with diameter $d$ and connected at the bottom with a Teflon
tube (F8006, Flon Industry, Japan) to the gas-flow controller (LogMIX, Front,
Japan). Through the controller, "air" (nitrogen gas of density $\rho_{A}=1.29$
g/cm$^{3}$; A3125N, Kenis) was injected into the cell at the bottom at a fixed
flux $Q$, while the cell was held in the upright position. The depth of the
observed region in the granular layer was more than approximately 10 cm from
the air-granular interface so that the granular pressure approaches a constant
value due to Janssen's mechanism \cite{Duran1997}. The cell width $W$ is
either $W=80$ mm or $W=40$ mm. We used the following three types of beads: (1)
glass beads with an average diameter $d=113$ $\mu$m (Bz01, As One), (2) glass
beads with an average diameter $d=196$ $\mu$m (Bz02, As One), and (3) Alumina
beads with an average diameter $d=124$ $\mu$m (Taimei Chemicals). The density
of glass beads is $\rho_{G}=2.5$ g/cm$^{3}$ and that of alumina beads is
$\rho_{G}=3.9$ g/cm$^{3}$.

\subsubsection*{Flow Regimes}

\begin{figure}[ptb]
\begin{center}
\includegraphics[width=\linewidth]{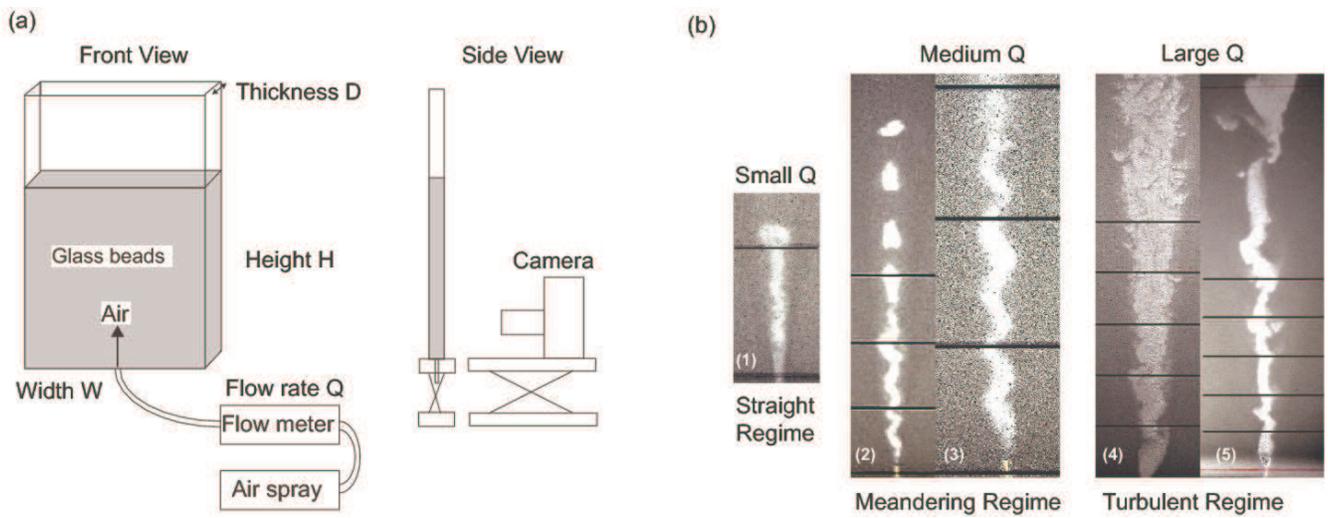}
\end{center}
\caption{(a) Experimental setup. (b) Flow-shape change with the flow rate $Q$.
Snapshots in (1)-(5) are obtained for the glass beads Bz01 for the cell width
$W=80$ mm under the conditions, $(D$ [mm], $Q$ [$\mu$m$^{3}/$s]$)=(1.0,0.5),$
$(1.0,1.25),$ $(1.0,1.17),$ $(0.7,1.67)$, and $(1.2,4.17)$.}%
\label{Fig1}%
\end{figure}

The flow created in the glass-bead bed is divided into three regimes, as shown
in Fig.~\ref{Fig1}(b) (movies for Bz01 are available as Supplementary Movie 1
- 3). (I) Straight regime: a nearly straight path ends in the middle of the
granular bed as a result of absorption of air by the bed. (II) Meandering
regime: regular wavy shapes appear in a transient but well-defined manner;
above the wavy path, bubbles are sometimes formed. (The observed meanders
appear near the outlet of the tube before they slightly travel along the path;
once stabilized the meander shapes do not travel.) (III) Turbulent regime:
different from the other regimes, the path goes through the bed to the
air-granular interface, and the flow appears as a hydrodynamic turbulent flow.
The turbulent flow tends to be straight for small $D$ (Fig.~\ref{Fig1}(b)-(4))
with fingering patterns along the side edges of the path, whereas the flow
tends to be wavy for large $D$ (Fig.~\ref{Fig1}(b)-(5)). Regimes I to III
appear in this order as the flow rate $Q$ increases for a given set of $D$,
$d$, and $\rho_{G}$.

\subsubsection*{Phase Diagram}

\begin{figure}[ptb]
\begin{center}
\includegraphics[width=0.8\linewidth]{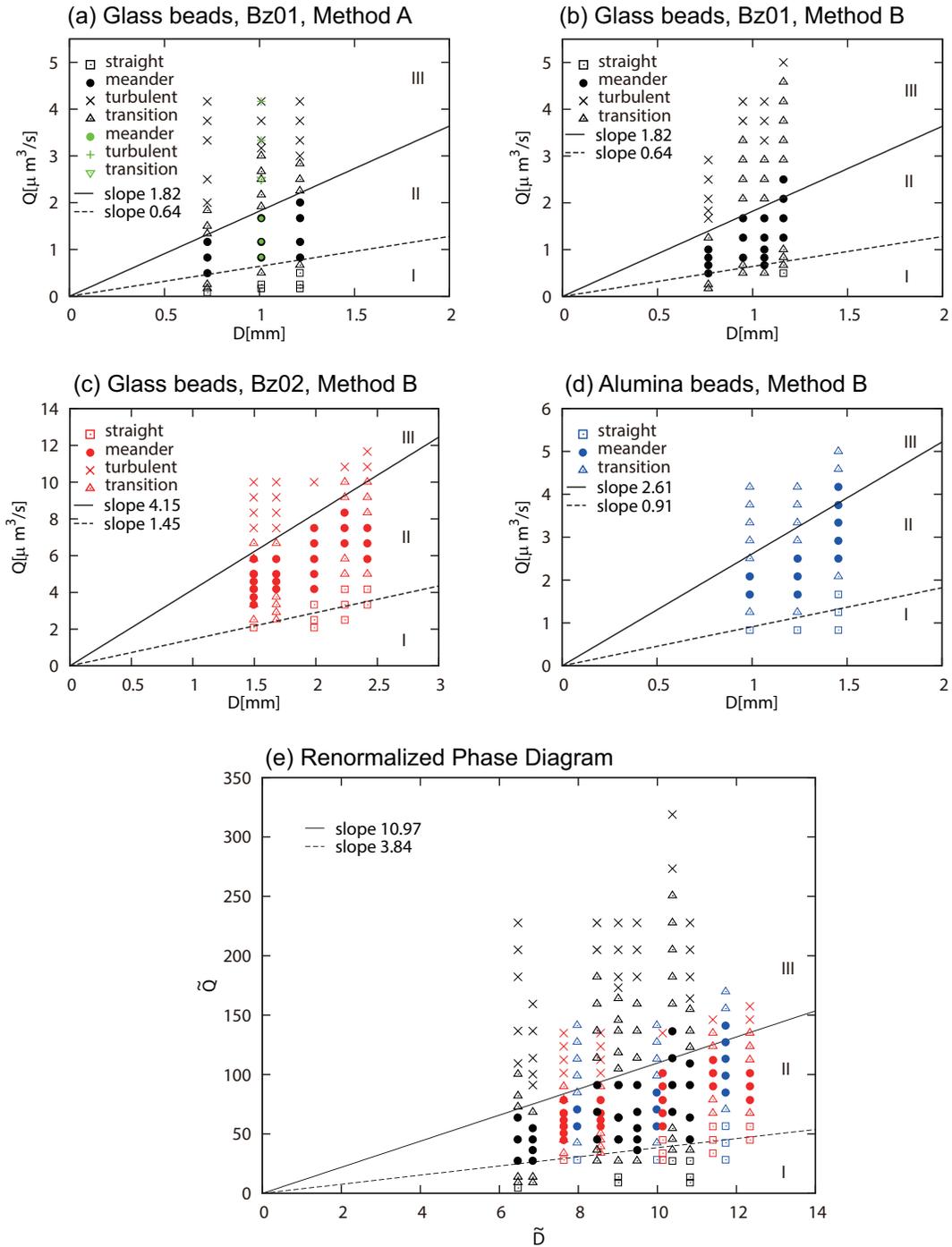}
\end{center}
\caption{(a)-(d) Phase diagrams, $D$ vs. $Q$, for different beads and packing
methods. The green symbols in (a) are the data obtained for $W=40$ mm. All of
the other data are obtained for $W=80$ mm. (e) Phase diagram with renormalized
axes. The labels, I, II, and III, represent the straight, meandering, and
turbulent regimes, respectively.}%
\label{Fig2}%
\end{figure}

In Fig.~\ref{Fig2}(a)-(d), the phase diagrams are shown on the $(D,Q)$ space.
The filled circles stand for the case in which the path is clearly in the
meandering regime (at least 4 waves can be recognized). The open squares
(crosses) stand for the case in which the path shape is clearly in the
straight (turbulent) regime. The open triangles represent the case in the
crossover region in which the waves are less than three and the path has
characteristics of more than two regimes. All the ensuing analyses of the
meandering path are performed for the data represented by the filed circles in
Fig.~\ref{Fig2}.

The two results for the glass beads, Bz01, shown in (a) and (b), demonstrate a
weak dependence on the way of packing the beads in the cell. We obtained (a)
through the following process (Method A): (1) we poured the beads in the cell,
(2) injected a strong gas flow of approximately $Q=5$ $\mu$m$^{3}/s$ and (3)
gradually decreased the flow to record the data points. In contrast, to obtain
(b), we added one more step (Method B): between step (1) and (2), we injected
a strong gas flow of approximately $Q=5$ $\mu$m$^{3}/s$ and gradually
decreased the flow to zero. In spite of this difference, the results in (a)
and (b) look reasonably similar to each other. In fact, the slight differences
in (a) and (b) may instead be due to differences in other conditions that
cannot be controlled precisely, such as humidity and electrostatic effects. In
the present study, all the data, except for the data in (a), are obtained
through Method B.

The phase diagrams in Fig.~\ref{Fig2}(a)-(d), obtained for glass beads, Bz01
and Bz02,\ and the alumina beads, look similar to one another. To quantify the
similarity, we collect in plot (e) all the data shown in (a)-(d), preserving
the colors and symbols in the phase diagram with renormalized axes, $D/d$ and
$Q/Q_{0}$ (the definition of $Q_{0}$ will be given later). The dashed and
solid lines in Fig.~\ref{Fig2}(e) are drawn according to the following
principles: we consider the slopes of the lines obtained by connecting the
origin with all the points represented by filled circles and select the line
with the smallest (largest) slope as the dashed (solid) line (in selecting the
solid line, we neglect two exceptional filled circles at $\tilde{D}=10.3$ and 11.7).

In Fig.~\ref{Fig2}(e), we clearly see universal or robust features of the
meandering phenomenon in the present study: the dashed and solid lines in
Fig.~\ref{Fig2}(e), which are selected as specified above, work reasonably
well as the lower and upper phase boundaries, respectively, irrespective of
$d$ and $\rho_{G}$. In fact, when the dashed and solid lines in (e) are mapped
back to the original plots, (a)-(d), the corresponding dashed and solid lines
in (a)-(d), work well as guides for the eyes to recognize the phase boundaries
in each plot. Note that the triangles corresponds to an intermediate state,
i.e., a path with characteristics of a meandering path. Further remarks will
be given in the Discussion.

In Fig.~\ref{Fig2}(a), it is shown that the change in the cell width $W$ has
practically no effect if the condition $D\ll W$ is satisfied. The symbols in
green are the data obtained with a cell of width $W=40$ mm. The green data
nearly overlap with the data obtained with a cell of width $W=80$ mm.

\subsubsection*{Self-similarity}

The centroid of the meandering shape is, in principle, of the form
$x=A\sin(2\pi y/\lambda)$, with amplitude $A$ and wave length $\lambda$, where
$x$ and $y$ are the vertical and horizontal positions within the cell,
respectively. If this description is essentially correct, the curvature $R$
will satisfy the scaling law%
\begin{equation}
1/R\simeq A/\lambda^{2} \label{e2}%
\end{equation}

As shown in Fig.~\ref{Fig3}(a)-(c), we found that the three characteristic
lengths all scale with the width $w$. From numerical fitting, we obtain%
\begin{equation}
\lambda=k_{1}w\text{, }A=k_{2}w,\text{ and }R=k_{3}w \label{e3}%
\end{equation}
with $k_{1}=3.31\pm0.13$, $k_{2}=0.445\pm0.02$ and $k_{3}=1.27\pm0.05$. These
relations are summarized as the following scaling laws:%
\begin{equation}
\lambda\simeq A\simeq R\simeq w, \label{eq0}%
\end{equation}
which means that the meandering shape is scale-invariant or self-similar:
there is a single length that characterizes all different meandering shapes.

\subsubsection*{Scaling law and renormalization}

\begin{figure}[ptb]
\begin{center}
\includegraphics[width=\linewidth]{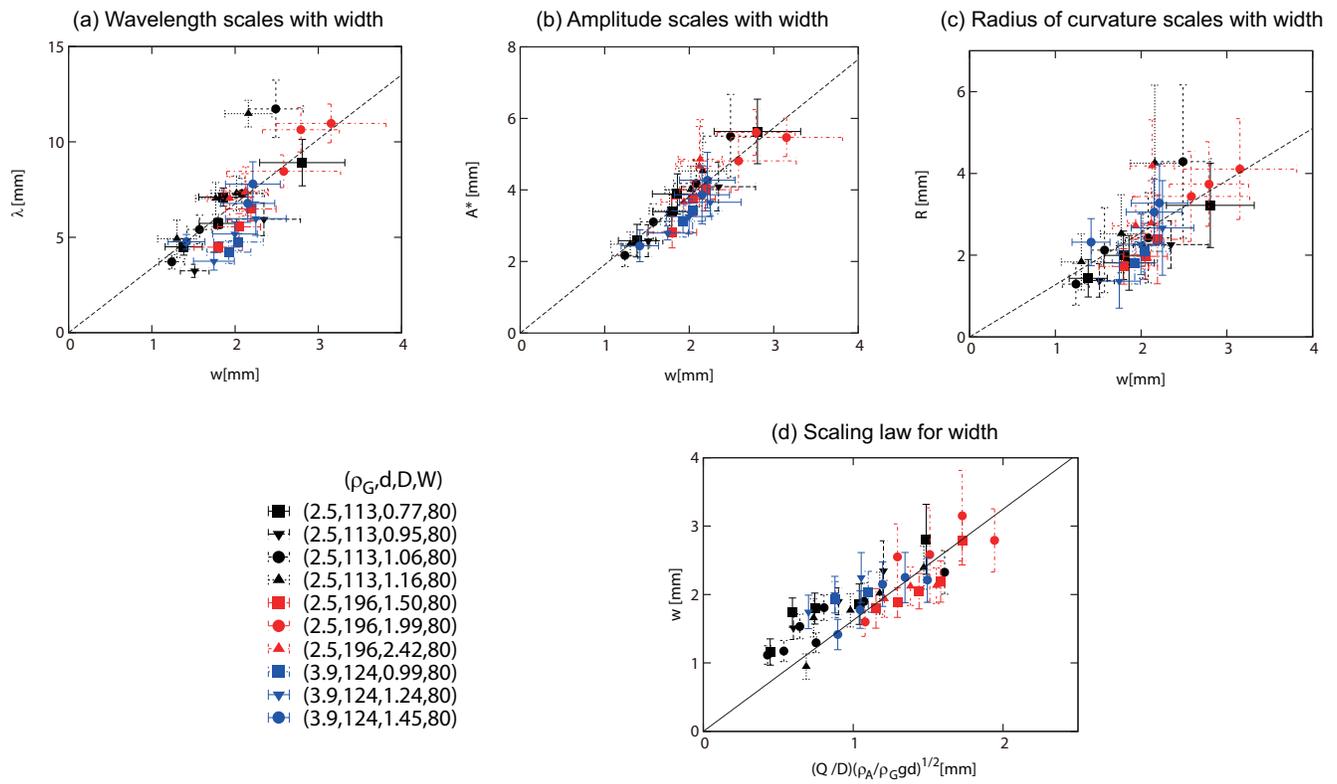}
\end{center}
\caption{(a)-(c) Wavelength $\lambda$, "amplitude" $A^{\ast}$, and radius of
curvature $R$ vs. width $w$, demonstrating that the three quantities are all
linearly dependent on $w.$ For the definition of $A^{\ast}$ see Fig.
\ref{Fig4}(b) and Methods. (d) $w$ vs. renormalized $Q$, demonstrating a
scaling law for $w$. The units of $\rho_{G},d,D$, and $W$ are g/cm$^{3}$,
$\mu$m, mm, and mm, respectively.}%
\label{Fig3}%
\end{figure}

As shown in Fig.~\ref{Fig3}(d), we experimentally found that the width $w$ is
well characterized by the following scaling law involving the gravitational
acceleration $g$:%
\begin{equation}
w\simeq(Q/D)\sqrt{\rho_{A}/(\rho_{G}gd)}. \label{eq1}%
\end{equation}
Here, as indicated in Fig.~\ref{Fig3}(d), the numerical coefficient is of the
order of unity, as expected. A simple physical explanation of this scaling law
will be given later.

The renormalization of the two axes performed for the phase diagram in
Fig.~\ref{Fig2}(e) is in fact motivated by this scaling law, which can be cast
into the form,%
\begin{equation}
\tilde{Q}\simeq(w/d)\tilde{D} \label{e5}%
\end{equation}
with $\tilde{Q}=Q/Q_{0}$ and $\tilde{D}=D/d$. The characteristic flow rate is
given by
\begin{equation}
Q_{0}=d^{2}V_{0}\text{ with }V_{0}=\sqrt{\rho_{G}gd/\rho_{A}} \label{e6}%
\end{equation}
Here, as will become clear shortly, $V_{0}$ scales with the velocity of the
air flow in the granular medium, which implies that $Q_{0}$ is the flow rate
for a section $\simeq d^{2}$. In other words, the renormalization of $D$ and
$Q$ employed in Fig.~\ref{Fig2}(e) corresponds to taking $d$ as the unit length.

Equation (\ref{e5}) suggests that the dashed and solid lines (phase
boundaries) in Fig.~\ref{Fig2}(e) correspond to\ the lower and upper bounds
for the width $w$ in the length unit $d$. This is because on the $\tilde{D}%
$-$\tilde{Q}$ plot, the data on a straight line going through the origin are
the collection of data having the same value of $w/d$. This implies that the
slopes of the dashed and solid lines correspond to the minimum and maximum
values of $w$ in the unit $d$.

\subsubsection*{Fluid-Solid Duality}

As shown in Fig.~\ref{Fig4}(a) the granular medium surrounding the meandering
path is categorized into fluidized and solidified regions. This snapshot is
taken with a relatively long exposure time, 1/20 sec, so that we can recognize
regions in which particles are almost at rest during this exposure time
(solidified region) and regions of the opposite character in which particles
are moving during the same exposure time (fluidized region). Quite naturally,
solidified regions are found near "convex" interfaces ("convex" when seen from
the side of the air path) because the air must be pushed back by the interface
to change its direction of flow (as a reaction, the air flow pushes the
interface through a centrifugal force). Fluidized regions are found near
concave interfaces, which is also natural because of the absence of
centrifugal force near the convex interface that, if it existed, would push
the interface. In addition, it is potentially noticeable in the snapshot that
the density in the fluidized region is lower. This suggests that particles
lose contacts with one another in the region, that is, the medium is
fluidized. These points are more clearly visible in a movie taken by a
high-speed camera (see Supplementary Movie 3). Note that the area of the
fluidized region is rather limited because of the principle of minimum
dissipation discussed below (in the ideal case, the area of the fluidized
region would be zero within the crudest approximation). Because of this, the
duality is not necessarily easy to recognize in Fig.~\ref{Fig4}(a) but is far
more clearly visualized in the movie, Supplementary Movie 3. The reader is
strongly encouraged to view this movie to confirm the distinction between the
solidified and fluidized regions.

\subsection*{Theory}

\subsubsection*{Principle of minimum dissipation}

\begin{figure}[ptb]
\begin{center}
\includegraphics[width=\linewidth]{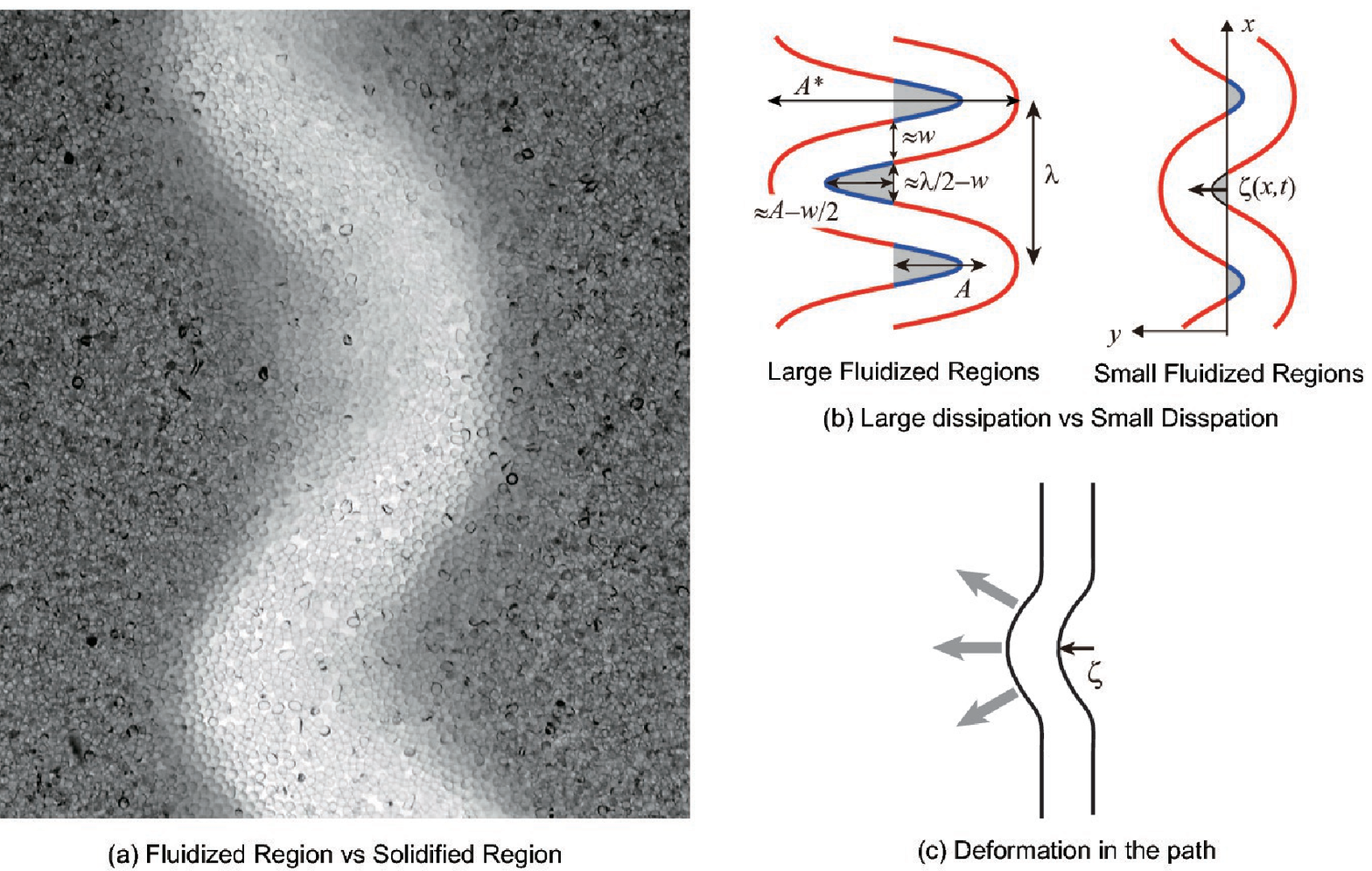}
\end{center}
\caption{(a) Visualization of fluidized and solidified regions via a snapshot
taken with a relatively long exposure time, in which the grains in the
fluidized region move but the grains in the solidified region do not. (b)
Illustration of wavy paths with large and small fluidized triangles. (c)
Illustration of a path with a small perturbation.}%
\label{Fig4}%
\end{figure}

The self-similarity of the meandering shape, described by the scaling laws in
Eq. (\ref{e3}), can be explained as a result of the principle of minimum
energy dissipation in the granular medium. Since the fluidized and solidified
regions are associated with the concave and convex parts, respectively, the
fluidized regions can be recognized as the triangular parts shown in gray in
Fig.~\ref{Fig4}(b). In these regions, energy is predominantly dissipated due
to inelastic collisions between particles: the area of a triangular region is
a measure of energy dissipation. The area scales as $(A-w/2)(\lambda/2-w)/2$,
as illustrated in \ref{Fig4}(b). This decreases to zero as the set
$(A,\lambda)$ approaches the following value:
\begin{equation}
(A,\lambda)\simeq(w/2,2w). \label{e8}%
\end{equation}
In other words, within this crude approximation, the area of the fluidized
region is optimized to zero. Accordingly, the fluidized region is in fact very
limited. (This is the reason the distinction between the solidified and
fluidized regions are not necessarily easy to recognize in a still snapshot,
but easy to recognize in a movie, as mentioned above.)

In fact, by inspection, we confirm that all three snapshots of meandering
paths in Fig.~\ref{Fig4}(a) and Fig.~\ref{Fig1}(b)-(2) \& (3) nearly satisfy
this condition! No paths resemble the path shown in the left panel of
Fig.~\ref{Fig4}(b) with large triangles; rather the path resemble the path
shown in the right panel of Fig.~\ref{Fig4}(b), with small triangles whose
areas are almost zero. At a quantitative level, we can confirm that
Eq.~(\ref{e8}) implies $k_{1}\simeq2$ and $k_{2}\simeq1/2$, which compare
quite well with the experimentally observed values given below Eq.~(\ref{e3}).

Equation (\ref{e8}), which is realized reasonably well in the experiments,
expresses that the dissipation is minimized when the scaling relation
$A\simeq\lambda\simeq w$ holds. These scaling laws are combined with Eq.
(\ref{e2}) to obtain the relation $R\simeq w$. Therefore, we confirm Eq.
(\ref{eq0}) on the basis of the principle of minimum dissipation.

\subsubsection*{Principle of the floating ping-pong ball}

The width $w$ of the meandering path is essentially determined by the
principle of the floating ping-pong ball \cite{FloatPingPong}, a familiar
phenomenon in which a ping-pong ball is levitated with air flow from a blower,
which is explained by Bernoulli's Principle. In the present case, at the top
end of the air path, as shown in the snapshot in Fig.~\ref{Fig1}(b)-(1) \&
(2), the sand particles are floating, similarly to a floating ping-pong ball.
At the level of scaling laws, this condition is expressed as the balance
between the dynamic pressure of the air flow of velocity $V$ and the
gravitational force acting on a sand particle, i.e., $\rho_{A}V^{2}\simeq
\rho_{G}gd^{3}/d^{2}$. This relation, which in fact means that $V$ scales as
$V_{0}$, which was already defined in Eq. (\ref{e6}), is combined with the
condition of the flow conservation,%
\begin{equation}
Q\simeq wDV, \label{e4}%
\end{equation}
to reveal the scaling law for the width given in Eq. (\ref{eq1}).

As shown in Fig.~\ref{Fig3}(d), the scaling law is reasonably well satisfied
for the meandering paths. This implies that, in the present experimental
conditions, the permeation of air in the granular medium is not significant at
the level of scaling laws. This is because the scaling law in Eq. (\ref{eq1}),
which is confirmed in Fig.~\ref{Fig3}(d), is derived with the assumption of
the flow conservation given in Eq. (\ref{e4}).

\subsubsection*{Minimal model for fluid-solid duality and linear instability}

The observed meandering instability can be understood on the basis of the
two-dimensional Navier-Stokes equation for the air flow. Since the Reynolds
number of the flow is relatively high, the distribution of the flow speed on
the section perpendicular to the flow is almost homogeneous, except near the
boundaries (i.e., the flow is a quasi plug flow). As a result, the dynamics
are well described by considering a path of linear density $\rho_{A}wD$
flowing with velocity $\mathbf{u}(x,y,t)=(u,v)$, with air density $\rho_{A}$
\cite{GondretHelleShawFlow,HinchHelleShawFlow,LimatPRL2011}.

Equation (\ref{eq1}), which has already been explained in terms of the
principle of the floating ping-pong ball, can also be explained in this
framework. When the path is nearly straight, the dominant component of the
velocity is $u$ ($u\gg v$) and its stationary dynamics is described simply by
$\rho_{A}uu_{x}+p_{x}=0$ (the subscripts denote partial derivatives and $p$ is
the isotropic pressure) because viscosity and gravity for the air flow can be
neglected; the quantity $\rho_{A}u^{2}/2+p$ is preserved along the path
(Bernoulli's principle). Throughout most of the path, this quantity scales as
$\rho_{A}V^{2}/2+p_{0}$, where $V$ is the velocity along the path and $p_{0}$
is the atmospheric pressure. However, it scales as $p_{G}+p_{0}$ near the top
end of the path: at the top of the path, as suggested above, granular
particles are fluidized with a high packing fraction, giving a pressure scale
$p_{G}$ of the order of $\rho_{G}gd^{3}/d^{2}$. Balancing the two pressure
scales, we obtain Eq. (\ref{eq1}).

The initiation of the meandering instability can be understood from the $y$
component of the equation of motion for the path of linear density $\rho
_{A}wD$, where a deformed path shape is described by $y=\zeta(x,t)$ (see
Fig.~\ref{Fig4}(c)). This equation of motion may be described by the equation
that minimally reflects the fluid-solid duality of granular materials along
the meandering path as illustrated in Fig.~\ref{Fig4}(a)-(b):
\begin{equation}
\rho_{A}wD(v_{t}+Vv_{x})=-\rho_{G}gD(K_{1}\zeta+K_{2}\tau\zeta_{t})
\label{eq12}%
\end{equation}
Here, $V$ is the air-flow velocity along the path. The left-hand side of this
equation stands for inertia, while the right-hand side expresses a simple and
minimal interaction of the flow with the granular medium for the instability:
the $K_{1}$ term expresses an elastic response, characterizing the solid-like
property, whereas the $K_{2}$ term represents a viscous response,
characterizing the liquid-like property, with the subscript $t$ denoting the
time derivative. Here, $K_{1}$ and $K_{2}$ are numerical coefficients of the
order of unity, whereas $\tau$ is a characteristic time.

\vspace{6pt}
\noindent\textit{Maximum wavelength for instability}

The linear stability analysis of Eq.~(\ref{eq12}) leads to two conditions for
the instability, with the one determining the maximum wavelength for the
instability. By seeking the solution to a linearized version of
Eq.~(\ref{eq12}) of the form $\zeta\simeq e^{\sigma t+iqx}$ with
$q=2\pi/\lambda$, we find that the solution becomes unstable (Re$[\sigma]>0$)
when the following two instability conditions are both satisfied:
\begin{equation}
\text{(1) }K_{2}\neq0\text{ and (2) }q^{2}V^{2}>\rho_{G}gK_{1}/(\rho_{A}w).
\label{eq4}%
\end{equation}
The first condition implies that the instability requires the $K_{2}$ term.
The second condition can be expressed as $\rho_{A}wV^{2}\zeta_{xx}>\rho
_{G}gK_{1}\zeta$; the centrifugal force overcoming the restoring force leads
to the instability. This is considerably different from the meandering
instability of rivulets interacting with solid plates \cite{LimatPRL2011}, in
which the centrifugal force does not play a role in triggering the
instability. The second condition also defines the maximum wavelength
$\lambda_{m}$ for unstable modes:
\begin{equation}
\lambda<\lambda_{m}\text{ \ with \ }\lambda_{m}=(2\pi)^{2}d/K_{1}.
\end{equation}

\vspace{6pt}
\noindent\textit{Fastest growing mode}

The scaling law $\lambda\simeq w$ in Eq. (\ref{eq0}), which has already been
justified on the basis of the principle of minimum dissipation, can also be
explained in the present framework: this relation emerges as a result of the
selection of the fastest-growing mode of the linear instability. The growth
velocity of the instability, measured by the quantity Re$[\sigma]$ ($>0$), is
a monotonically increasing function of $q^{2}$; the smaller the wavelength,
the faster the growth of the instability. However, there exists a minimum
wavelength in the present theory; the condition $\lambda>w$ should be
satisfied, that is, the wavelength of the fastest-growing mode is practically
given by $\lambda\simeq w$. This is because the present theory is valid only
when the width of the path is smaller than any other length scales
characterizing the shape of the path; the flow is here treated as a (curved)
"line." In this way, the wavelength of the fastest-growing mode $\lambda\simeq
w$ tends to be selected as the wavelength of the meandering path.

\vspace{6pt}
\noindent\textit{Phase boundaries}

The existence of minimum and maximum values for $w$ in the length unit $d$,
suggested in the renormalized phase diagram in shown Fig.~\ref{Fig2}(e), can
be explained in the present framework. The maximum value of $w$ is given as
$\lambda_{m}$ from the relations, $\lambda\simeq w$ and $\lambda<\lambda
_{m}=(2\pi)^{2}d/K_{1}$, which were already justified. The minimum of $w$ is
given by the limitation of the present continuum description: the path
interface can be regarded as smooth only when $w$ is significantly larger than
$d$, that is, the condition $kd<w$ must be satisfied for the continuum theory
with $k$ larger than unity. In summary, we expect
\begin{equation}
k<w/d<(2\pi)^{2}/K_{1}.
\end{equation}
The minimum and maximum of $w/d$, $k$ and $(2\pi)^{2}/K_{1}$, should
correspond, respectively, to the slopes of the dashed and solid lines (phase
boundaries) in Fig.~\ref{Fig2}(e), which are 3.84 and 11.0, respectively. From
this, we obtain $k\simeq3.84$ and $K_{1}\simeq3.59$, the orders of magnitude
of which are consistent with the scaling arguments given above.

This analysis suggests that the solid-like interaction of the form $\rho
_{G}gDK_{1}\zeta$ employed in Eq.~(\ref{eq12}) is rather universal and does
not depend on the stiffness of the particles. As shown in Fig.~\ref{Fig2}, the
phase boundaries are rather universal for the glass and alumina beads, whereas
Young's modulus of alumina is about five times as large as that of glass. This
implies that the value of $K_{1}$ is almost the same for the glass and alumina
beads despite the large difference in the particle-level stiffness. This may
be because the interaction between the air flow and "granular solid" is weak
in the sense that it is not related to the deformation associated with the
Hertz contact \cite{Landau} but rather to the gravity acting on granular
materials (note that the term is proportional to $\rho_{G}g$).

\section*{Discussion}

Based on our experiment and theory, we have elucidated physical origins of the
meandering instability and scale invariance by considering the interaction of
the flow with the granular medium, which exhibits dual characters of solid and
liquid. (1) Meandering instability occurs for an air flow whose width is fixed
by the principle of the floating ping-pong ball. (2) This flow is destabilized
due to a linear instability that results from the competition between the
centrifugal force (inducing instability) and the restoring force associated
with the jammed or solidified granular medium (inducing stability); the flow
selects the fastest-growing mode whose wavelength is comparable to the width.
(3) The growth of the amplitude of the mode with the selected wavelength is
suspended as a result of the competition between the linear instability and
the energy dissipation taking place in the granular medium, which is brought
about by the fluidized granular medium. The instability tends to increase the
amplitude of the path, but when the amplitude becomes larger, the dissipation
becomes larger, and the velocity is reduced, thus mitigating the centrifugal
force, which is the source of the instability. This subtle balance emerges as
the principle of minimum dissipation, which sets the quasi-static amplitude
comparable to the width. These mechanisms justify the scale invariance and the
phase boundary lines for the meandering regime.

River meandering is a well-known phenomenon and is related in a number of ways
to the work described in this paper. Compared with river meandering, the time
scale of the meandering instability studied here is extremely small. The
present meandering paths frequently disappears, which makes them quasi-static.
(This may be due to noise in the flow, possibly created at the exit of the
tube; such nose could trigger the instability on its own). Even deep in the
meandering regime, the meandering shape is only stable on the sub-second time
scale, as suggested in Fig.~\ref{Fig4}(a) and Supplementary Movies 1-3. In
addition, the erosion and accumulation processes, which are crucial to
river-meander formation, are considerably different from the processes
associated with fluid-solid duality, which is essential to the meandering
phenomenon considered in this paper. Nonetheless, the two meandering
instabilities share several similarities. Meandering fluids in both cases
interact not with an undeformable solid but with a deformable "complex fluid."
In addition, the self-similarity property in Eq. (\ref{e3}) has also been
established through field studies of river meandering, and this property seems
to be universal and applicable even to meltwater streams on the surfaces of
glaciers \cite{MeanderLeopold}. Remarkably, the numerical coefficients,
$k_{1}$ through $k_{3}$, found for river meandering are comparable to those
found here. We thus expect that the arguments developed here would be useful
as a starting point for a physical understanding of river meandering.

This study is also relevant to fluidized beds. Fluidized granular beds have
been extensively studied experimentally
\cite{davidson1963fluidised,davidson1985fluidization,FluidizeExpRev1994} and
numerically \cite{NumericalFluidized1996Swaaij,NumericalFluidized1997}.
However, in most cases, air is homogeneously injected at the base of the
container, which is quite different from the local injection in the present
study. In these homogeneous cases, bubbling has been thoroughly studied.
(However, note that even in the less-studied cases in which jets are injected,
no attention has been paid to meandering instability
\cite{davidson1985fluidization,FluidizedJet2005,ENSMeanderGranular2011,FluidizedJet2015}%
.) For example, it is known that control of the dynamics of the bubbles
created in fluidized reactors could help enhance the efficiency of the
reactors \cite{BubbleEffectsFluidizedBed1981,FluidizeExpRev1994}. In the
context of fluidized granular beds, the effect of air permeation is very
important. The viscous force originating from the interstitial air in the
porous medium governs the onset of the bubbling in the bed when the air
injection is homogeneous. In contrast, this effect is minor in the present
case. This is strongly supported by Figs. \ref{Fig2} and \ref{Fig3}(c): the
discussions associated with the figures relay on Eq. (\ref{eq1}), in the
derivation of which the flow conservation given in Eq. (\ref{e4}) is assumed.
Fluidization processes are essential for industrial applications, such as
temperature control, heat transfer, coal combustion and fluidized bed
reactors, which are useful for reactions in the conversion of crude oils to
gasoline and biomass gasification. The present study is relevant to fluidized
beds and might be of some use in such industrial applications.

To compare the present results with those obtained in other contexts, such as
geoscience and fluidized beds, it would be useful to discuss these results in
terms of more generic dimensionless factors that are introduced from physical
pictures at the level of a single particle. (i) In geophysics, the Shields
number is quite often used\cite{ShieldsNumPascale2007}. This number compares
(for a grain) the total shear force with the gravitational force, and has been
used to discuss the onset of sedimentation. In the present phenomena, a
similar number could be defined through the floating condition for a grain (or
a \textquotedblleft ping-pong ball\textquotedblright), i.e., the relative
importance of the force due to the dynamic pressure compared with the
gravitational force. In the present study, this number is a constant of the
order of unity in the meandering regime (see the paragraph leading to Eq.
(8)). (ii) In the previous study in Ref.\cite{SinuosityPNAS2013}, it is shown
that the channel sinuosity, defined as the ratio of the total length along the
path to the length projected onto the direction of the average flow, is weakly
dependent on the Froude number, which can be regarded as the ratio between the
gradient of altitude for a channel in the average-flow direction and the
resistance to the flow due to vegetation. The sinuosity values obtained from
20 rivers around the world range from 1.2 to 2.2. In the present case, the
sinuosity scales as $R/\lambda$, i.e., the sinuosity is a constant of the
order of unity, which is not in conflict with this previous result and is in
accordance with the result of the study in Ref.\cite{MeanderLeopold}. (iii) In
terms of the Reynolds number for a grain, which is defined as the ratio of the
inertial force to the viscous force acting on a grain, in the present study,
the fluidized parts of grains are important to the discussion of the dynamic
behaviors, and such parts are in the inertial regime, i.e., the Reynolds
number is larger than unity, because the interstitial fluid is air in this
experiment. However, for the discussion on the solidified part, it is possible
that the viscosity of the interstitial air could play a role. (iv) In granular
physics, the inertial number has been established to describe a constitutive
law for dry granular materials\cite{midi2004dense} and the idea is extended to
cases with interaction with interstitial
fluid\cite{cassar2005submarine,du2003granular}. The same constitutive law
could hold for dry and immersed granular materials if the meaning of the
inertial number is properly modified. The inertial number for a dry granular
material can be regarded as the ratio between an inertial time for
rearrangement due to a pressure and a macroscopic time spent by the particle
to move from one hole to the next. For the present air flows in a granular
medium, the counterpart for the latter time is $d/V$. The counterpart for the
former rearrangement time may be a time associated with the dynamic pressure
or with the gravitational force, and this time scale is given by $d\sqrt
{\rho_{G}/P}$ with $P\simeq\rho_{A}V^{2}$ or $P\simeq\rho_{G}gd$, i.e.,
$t_{d}\simeq(V/d)\sqrt{\rho_{G}/\rho_{A}}$ or $t_{g}\simeq\sqrt{d/g}$,
respectively. These times, $t_{d}$ and $t_{g},$ are of the same order of
magnitude in the present case because of the floating condition. The single
time scale for rearrangements, $t_{d}\simeq t_{g}$, is typically a few ms,
which is consistent with the observation made with a high-speed camera (e.g.,
Supplementary Movie 3). By comparing this rearrangement time scale with the
macroscopic time scale, we define the dimensionless number $\sqrt{\rho
_{G}/\rho_{A}}$, as introduced in the previous studies
\cite{cassar2005submarine,du2003granular}. This number would also be useful in
comparing the present results with those obtained in the other contexts.

\section*{Conclusion}

In this study, we show that an air flow in a thin granular bed can be
destabilized to show meandering shape. The phase diagram for the meandering
regime is shown in a universal way as a function of a normalized flow rate and
a normalized granular bed thickness. In addition, we show that the meandering
shapes are self-similar or scale invariant, as observed in river meandering.
These experimental results lead to physical insights as summarized in the
first paragraph of Discussion.

Fundamentally, our results open a new avenue in the field of granular physics
\cite{Duran1997,PouliquenBook,HerminghausBook} by proposing a minimal
description for the fluid-solid duality and a principle of minimal
dissipation, leading to a new opportunity for fruitful connections of granular
physics to various fields. For example, the physical description in
Eq.~(\ref{eq12}) of the interplay between jamming
\cite{LiuNagel,TakeharaPRL2014} (solid-like response) and fluidization
\cite{BehringerNature94Fluidized} (dissipation) will be useful in
understanding soil-bed fluidization induced by earthquakes
\cite{howard1989fluidized}, and the present framework, including the principle
of minimum dissipation, may help geophysicists to physically understand river
meandering. Practically, the present study could be useful for industrial
issues such as fluidized-bed reactors.

\section*{Methods}

\subsection*{Experimental}

The experiment was recorded using a digital camera (D800E, Nikon, Japan) to
obtain the data for analysis. For high-speed visualization of the solidified
and fluidized regions, a high-speed camera (UX100, Photoron, Japan) was also
used. To both cameras, a macro lens (Micro NIKKOR 60 mm F2.8 ED, Nikon, Japan)
was attached. The cell thickness $D$ was measured by a laser distance sensor
(ZS-HLDS5+ZS-HLDC11+Smart Monitor Zero Pro., Omron, Japan).

\subsection*{Data Analysis}

The measurements of $w$, $A$, $\lambda$, and $R$ are performed as follows. The
average value and the error bar (the standard deviation) of $w$ and $R$ for a
single data point in Fig.~\ref{Fig3}(a)-(c) are obtained from 30 to 60
measurements. We used 10 to 20 snapshots acquired under the same conditions
(the same $\rho_{G}$, $d$, $D$, and $W$) and selected three points
(corresponding to $y=n\lambda$ with $n$ an integer) from each snapshot to
perform the measurements. In each measurement, $R$ is determined by fitting a
parabolic form to the outside edge of a path. As for $\lambda$ and $A^{\ast}$
(see Fig.~\ref{Fig4}(b)), we measured 10 to 20 times for a fixed condition by
using 10 to 20 snapshots and making one measurement for each snapshot. In each
measurement, $\lambda$ and $A^{\ast}$ are determined by selecting a segment
containing $n$ waves (with $n$ an integer equal to or larger than 4). To
estimate $\lambda$ the length of the segment in the $x$ direction is measured,
and then divided by the number of waves $n$. To obtain $A^{\ast}$, $A_{\min
}^{\ast}$ and $A_{\max}^{\ast}$ are determined, and then we take the average
of these two quantities (We measured $A^{\ast}$ rather than $A$ because the
former quantity can be measured in a less ambiguous manner, and then used the
relation $A^{\ast}=2(A+w/2)$ to obtain $A$).

\subsection*{Theory}

Details of the linear stability analysis are outlined as follows.
Equation~(\ref{eq12}), combined with $v=\zeta_{t}+V\zeta_{x}$, leads to the
equation, $\zeta_{tt}+2V\zeta_{xt}+V^{2}\zeta_{xx}+\kappa_{1}\zeta+\kappa
_{2}\zeta_{t}=0$, linearized in $\zeta$, where $\kappa_{1}=\rho_{G}%
gK_{1}/(\rho_{A}w)$ and $\kappa_{2}=\rho_{G}gK_{2}\tau/(\rho_{A}w)$. The
solution of the form $e^{\sigma t+iqx}$ with $q=2\pi/\lambda$ satisfies
$\sigma=-iqV-\kappa_{2}/2\pm\sqrt{(\kappa_{2}/2)^{2}-\kappa_{1}+iqV\kappa_{2}%
}$. The condition for meandering instability $\operatorname{Re}\sigma>0$
results in the conditions given in Eq. (\ref{eq4}).


\begin{thebibliography}{99}                                                                                               %
\expandafter\ifx\csname url\endcsname\relax


\fi
\expandafter\ifx\csname urlprefix\endcsname\relax


\fi
\providecommand{\bibinfo}[2]{#2} \providecommand{\eprint}[2][]{\url{#2}}

\bibitem {CosmicHelixScience2001}\bibinfo{author}{Lobanov, A.} \&
\bibinfo{author}{Zensus, J.}
\newblock \bibinfo{title}{A cosmic double helix in the archetypical quasar
3c273}. \newblock \emph{\bibinfo{journal}{Science}}
\textbf{\bibinfo{volume}{294}}, \bibinfo{pages}{128--131} (\bibinfo{year}{2001}).

\bibitem {MantleTransportNature1988}\bibinfo{author}{Whitehead, J.} \&
\bibinfo{author}{Helfrich, K.}
\newblock \bibinfo{title}{Wave transport of deep mantle material}.
\newblock \emph{\bibinfo{journal}{Nature}} \textbf{\bibinfo{volume}{336}},
\bibinfo{pages}{59--61} (\bibinfo{year}{1988}).

\bibitem {kessler1986JFM86}\bibinfo{author}{Kessler, J.~O.}
\newblock \bibinfo{title}{Individual and collective fluid dynamics of swimming
cells}. \newblock \emph{\bibinfo{journal}{J. Fluid. Mech.}}
\textbf{\bibinfo{volume}{173}}, \bibinfo{pages}{191--205} (\bibinfo{year}{1986}).

\bibitem {CoilLiquidPRLGoldstein2005}\bibinfo{author}{Dombrowski, C.}
\emph{et~al.}
\newblock \bibinfo{title}{Coiling, entrainment, and hydrodynamic coupling of
decelerated fluid jets}. \newblock \emph{\bibinfo{journal}{Phys. Rev. Lett.}}
\textbf{\bibinfo{volume}{95}}, \bibinfo{pages}{184501} (\bibinfo{year}{2005}).

\bibitem {Nakagawa1984}\bibinfo{author}{Nakagawa, T.} \&
\bibinfo{author}{Scott, J.~C.}
\newblock \bibinfo{title}{Stream meanders on a smooth hydrophobic surface}.
\newblock \emph{\bibinfo{journal}{J. Fluid Mech.}}
\textbf{\bibinfo{volume}{149}}, \bibinfo{pages}{89--99} (\bibinfo{year}{1984}).

\bibitem {braiding}\bibinfo{author}{Mertens, K.},
\bibinfo{author}{Putkaradze, V.} \& \bibinfo{author}{Vorobieff, P.}
\newblock \bibinfo{title}{Braiding patterns on an inclined plane}.
\newblock \emph{\bibinfo{journal}{Nature}} \textbf{\bibinfo{volume}{430}},
\bibinfo{pages}{165--165} (\bibinfo{year}{2004}).

\bibitem {Drenckhan2004}\bibinfo{author}{Drenckhan, W.},
\bibinfo{author}{Gatz, S.} \& \bibinfo{author}{Weaire, D.}
\newblock \bibinfo{title}{Wave patterns of a rivulet of surfactant solution in
a hele-shaw cell}. \newblock \emph{\bibinfo{journal}{Phys. Fluids}}
\textbf{\bibinfo{volume}{16}}, \bibinfo{pages}{3115} (\bibinfo{year}{2004}).

\bibitem {Braudrick2009}\bibinfo{author}{Braudrick, C.~A.},
\bibinfo{author}{Dietrich, W.~E.}, \bibinfo{author}{Leverich, G.~T.} \&
\bibinfo{author}{Sklar, L.~S.}
\newblock \bibinfo{title}{Experimental evidence for the conditions necessary to
sustain meandering in coarse-bedded rivers}.
\newblock \emph{\bibinfo{journal}{Proc. Nat. Acad. Sci. USA}}
\textbf{\bibinfo{volume}{106}}, \bibinfo{pages}{16936--16941} (\bibinfo{year}{2009}).

\bibitem {LimatPRL06}\bibinfo{author}{Le~Grand-Piteira, N.},
\bibinfo{author}{Daerr, A.} \& \bibinfo{author}{Limat, L.}
\newblock \bibinfo{title}{Meandering rivulets on a plane: A simple balance
between inertia and capillarity}.
\newblock \emph{\bibinfo{journal}{Phys. Rev. Lett.}}
\textbf{\bibinfo{volume}{96}}, \bibinfo{pages}{254503} (\bibinfo{year}{2006}).

\bibitem {LimatPRL2011}\bibinfo{author}{Daerr, A.},
\bibinfo{author}{Eggers, J.}, \bibinfo{author}{Limat, L.} \&
\bibinfo{author}{Valade, N.}
\newblock \bibinfo{title}{General mechanism for the meandering instability of
rivulets of newtonian fluids}.
\newblock \emph{\bibinfo{journal}{Phys. Rev. Lett.}}
\textbf{\bibinfo{volume}{106}}, \bibinfo{pages}{184501} (\bibinfo{year}{2011}).

\bibitem {DaerrEPL2012}\bibinfo{author}{Couvreur, S.} \&
\bibinfo{author}{Daerr, A.}
\newblock \bibinfo{title}{The role of wetting heterogeneities in the meandering
instability of a partial wetting rivulet}.
\newblock \emph{\bibinfo{journal}{EPL (Europhys. Lett.)}}
\textbf{\bibinfo{volume}{99}}, \bibinfo{pages}{24004} (\bibinfo{year}{2012}).

\bibitem {Birnir2008}\bibinfo{author}{Birnir, B.},
\bibinfo{author}{Mertens, K.}, \bibinfo{author}{Putkaradze, V.} \&
\bibinfo{author}{Vorobieff, P.}
\newblock \bibinfo{title}{Meandering fluid streams in the presence of flow-rate
fluctuations}. \newblock \emph{\bibinfo{journal}{Phys. Rev. Lett.}}
\textbf{\bibinfo{volume}{101}}, \bibinfo{pages}{114501} (\bibinfo{year}{2008}).

\bibitem {Howard2009}\bibinfo{author}{Howard, A.~D.}
\newblock \bibinfo{title}{How to make a meandering river}.
\newblock \emph{\bibinfo{journal}{Proc. Nat. Acad. Sci. USA}}
\textbf{\bibinfo{volume}{106}}, \bibinfo{pages}{17245--17246} (\bibinfo{year}{2009}).

\bibitem {Lewin1972}\bibinfo{author}{Lewin, J.}
\newblock \bibinfo{title}{Late-stage meander growth}.
\newblock \emph{\bibinfo{journal}{Nature}} \textbf{\bibinfo{volume}{240}},
\bibinfo{pages}{116--116} (\bibinfo{year}{1972}).

\bibitem {Trinkaus1998}\bibinfo{author}{Trinkaus, J.}
\newblock \bibinfo{title}{Gradient in convergent cell movement during
fundulusgastrulation}.
\newblock \emph{\bibinfo{journal}{J. Experiment. Zoology}}
\textbf{\bibinfo{volume}{281}}, \bibinfo{pages}{328--335} (\bibinfo{year}{1998}).

\bibitem {Napieralski1996}\bibinfo{author}{Napieralski, J.~A.} \&
\bibinfo{author}{Eisenman, L.~M.}
\newblock \bibinfo{title}{Further evidence for a unique developmental
compartment in the cerebellum of the meander tail mutant mouse as revealed by
the quantitative analysis of purkinje cells}.
\newblock \emph{\bibinfo{journal}{J. Comparative Neurology}}
\textbf{\bibinfo{volume}{364}}, \bibinfo{pages}{718--728} (\bibinfo{year}{1996}).

\bibitem {Flierl1993}\bibinfo{author}{Flierl, G.~R.} \&
\bibinfo{author}{Davis, C.~S.}
\newblock \bibinfo{title}{Biological effects of gulf stream meandering}.
\newblock \emph{\bibinfo{journal}{J. Marine Res.}}
\textbf{\bibinfo{volume}{51}}, \bibinfo{pages}{529--560} (\bibinfo{year}{1993}).

\bibitem {Sawai1981}\bibinfo{author}{Ikeda, S.}, \bibinfo{author}{Parker, G.}
\& \bibinfo{author}{Sawai, K.}
\newblock \bibinfo{title}{Bend theory of river meanders. part 1. linear
development.} \newblock \emph{\bibinfo{journal}{J. Fluid Mech.}}
\textbf{\bibinfo{volume}{112}}, \bibinfo{pages}{363--377} (\bibinfo{year}{1981}).

\bibitem {Howard1984}\bibinfo{author}{Howard, A.~D.} \&
\bibinfo{author}{Knutson, T.~R.}
\newblock \bibinfo{title}{Sufficient conditions for river meandering: A
simulation approach}. \newblock \emph{\bibinfo{journal}{Water Resour. Res.}}
\textbf{\bibinfo{volume}{20}}, \bibinfo{pages}{1659--1667} (\bibinfo{year}{1984}).

\bibitem {Stolum1996}\bibinfo{author}{Stolum, H.-H.}
\newblock \bibinfo{title}{River meandering as a self-organization process}.
\newblock \emph{\bibinfo{journal}{Science}} \textbf{\bibinfo{volume}{271}},
\bibinfo{pages}{1710--1713} (\bibinfo{year}{1996}).

\bibitem {SwinneyPRL96}\bibinfo{author}{Li, G.}, \bibinfo{author}{Ouyang, Q.},
\bibinfo{author}{Petrov, V.} \& \bibinfo{author}{Swinney, H.~L.}
\newblock \bibinfo{title}{Transition from simple rotating chemical spirals to
meandering and traveling spirals}.
\newblock \emph{\bibinfo{journal}{Phys. Rev. Lett.}}
\textbf{\bibinfo{volume}{77}}, \bibinfo{pages}{2105} (\bibinfo{year}{1996}).

\bibitem {Howard1970}\bibinfo{author}{Howard, A.~D.},
\bibinfo{author}{Keetch, M.~E.} \& \bibinfo{author}{Vincent, C.~L.}
\newblock \bibinfo{title}{Topological and geometrical properties of braided
streams}. \newblock \emph{\bibinfo{journal}{Water Ressour. Res.}}
\textbf{\bibinfo{volume}{6}}, \bibinfo{pages}{1674--1688} (\bibinfo{year}{1970}).

\bibitem {Ferguson1976}\bibinfo{author}{Ferguson, R.}
\newblock \bibinfo{title}{Disturbed periodic model for river meanders}.
\newblock \emph{\bibinfo{journal}{Earth Surf. Processes}}
\textbf{\bibinfo{volume}{1}}, \bibinfo{pages}{337--347} (\bibinfo{year}{1976}).

\bibitem {Bruinsma1990}\bibinfo{author}{Bruinsma, R.}
\newblock \bibinfo{title}{The statistical mechanics of meandering}.
\newblock \emph{\bibinfo{journal}{J. Phys. France}}
\textbf{\bibinfo{volume}{51}}, \bibinfo{pages}{829--845} (\bibinfo{year}{1990}).

\bibitem {EdwardsMeander}\bibinfo{author}{Liverpool, T.~B.} \&
\bibinfo{author}{Edwards, S.~F.}
\newblock \bibinfo{title}{Dynamics of a meandering river}.
\newblock \emph{\bibinfo{journal}{Phys. Rev. Lett.}}
\textbf{\bibinfo{volume}{75}}, \bibinfo{pages}{3016} (\bibinfo{year}{1995}).

\bibitem {MeanderLeopold}\bibinfo{author}{Leopold, L.~B.} \&
\bibinfo{author}{Wolman, M.~G.} \newblock \bibinfo{title}{River meanders}.
\newblock \emph{\bibinfo{journal}{Geol. Soc. Ame. Bull.}}
\textbf{\bibinfo{volume}{71}}, \bibinfo{pages}{769--793} (\bibinfo{year}{1960}).

\bibitem {SinuosityPNAS2013}\bibinfo{author}{Lazarus, E.~D.} \&
\bibinfo{author}{Constantine, J.~A.}
\newblock \bibinfo{title}{Generic theory for channel sinuosity}.
\newblock \emph{\bibinfo{journal}{Proc. Nat. Acad. Sci. (USA)}}
\textbf{\bibinfo{volume}{110}}, \bibinfo{pages}{8447--8452} (\bibinfo{year}{2013}).

\bibitem {Duran1997}\bibinfo{author}{Duran, J.}
\newblock \emph{\bibinfo{title}{Sables Poudres et Grains}}
(\bibinfo{publisher}{Editions Eyrolles in Paris}, \bibinfo{year}{1997}).

\bibitem {FloatPingPong}\bibinfo{author}{Wild, P.},
\bibinfo{author}{Surgenor, B.} \& \bibinfo{author}{Zak, G.}
\newblock \bibinfo{title}{The mechatronics laboratory experience}.
\newblock \emph{\bibinfo{journal}{Mechatronics}} \textbf{\bibinfo{volume}{12}}%
, \bibinfo{pages}{207--215} (\bibinfo{year}{2002}).

\bibitem {GondretHelleShawFlow}\bibinfo{author}{Gondret, P.} \&
\bibinfo{author}{Rabaud, M.}
\newblock \bibinfo{title}{Shear instability of two-fluid parallel flow in a
hele--shaw cell}. \newblock \emph{\bibinfo{journal}{Phys. Fluids}}
\textbf{\bibinfo{volume}{9}}, \bibinfo{pages}{3267--3274} (\bibinfo{year}{1997}).

\bibitem {HinchHelleShawFlow}\bibinfo{author}{Plourabou{\'e}, F.} \&
\bibinfo{author}{Hinch, E.~J.}
\newblock \bibinfo{title}{Kelvin--helmholtz instability in a hele-shaw cell}.
\newblock \emph{\bibinfo{journal}{Phys. Fluids}} \textbf{\bibinfo{volume}{14}}%
, \bibinfo{pages}{922--929} (\bibinfo{year}{2002}).

\bibitem {Landau}\bibinfo{author}{Landau, L.} \&
\bibinfo{author}{Lifshitz, E.}
\newblock \emph{\bibinfo{title}{Elasticity theory}}
(\bibinfo{publisher}{Pergamon Press}, \bibinfo{year}{1975}).

\bibitem {davidson1963fluidised}\bibinfo{author}{Davidson, J.~F.} \&
\bibinfo{author}{Harrison, D.}
\newblock \emph{\bibinfo{title}{Fluidised Particles}}
(\bibinfo{publisher}{Cambridge Univ. Press}, \bibinfo{year}{1963}).

\bibitem {davidson1985fluidization}\bibinfo{author}{Davidson, J.~F.},
\bibinfo{author}{Clift, R.} \& \bibinfo{author}{Harrison, D.}
\newblock \emph{\bibinfo{title}{Fluidization, 2nd. Ed.}}
(\bibinfo{publisher}{Academic Press, Inc., Orlando, FL}, \bibinfo{year}{1985}).

\bibitem {FluidizeExpRev1994}\bibinfo{author}{Yates, J.} \&
\bibinfo{author}{Simons, S.}
\newblock \bibinfo{title}{Experimental methods in fluidization research}.
\newblock \emph{\bibinfo{journal}{Int. J. Multiphase Flow}}
\textbf{\bibinfo{volume}{20}}, \bibinfo{pages}{297--330} (\bibinfo{year}{1994}).

\bibitem {NumericalFluidized1996Swaaij}\bibinfo{author}{Hoomans, B.},
\bibinfo{author}{Kuipers, J.}, \bibinfo{author}{Briels, W.} \&
\bibinfo{author}{Van~Swaaij, W.}
\newblock \bibinfo{title}{Discrete particle simulation of bubble and slug
formation in a two-dimensional gas-fluidised bed: a hard-sphere approach}.
\newblock \emph{\bibinfo{journal}{Chem. Eng. Sci.}}
\textbf{\bibinfo{volume}{51}}, \bibinfo{pages}{99--118} (\bibinfo{year}{1996}).

\bibitem {NumericalFluidized1997}\bibinfo{author}{Xu, B.} \&
\bibinfo{author}{Yu, A.}
\newblock \bibinfo{title}{Numerical simulation of the gas-solid flow in a
fluidized bed by combining discrete particle method with computational fluid
dynamics}. \newblock \emph{\bibinfo{journal}{Chem. Eng. Sci.}}
\textbf{\bibinfo{volume}{52}}, \bibinfo{pages}{2785--2809} (\bibinfo{year}{1997}).

\bibitem {FluidizedJet2005}\bibinfo{author}{Patil, D.},
\bibinfo{author}{van Sint~Annaland, M.} \& \bibinfo{author}{Kuipers, J.}
\newblock \bibinfo{title}{Critical comparison of hydrodynamic models for gas -
solid fluidized beds - part i: bubbling gas--solid fluidized beds operated
with a jet}. \newblock \emph{\bibinfo{journal}{Chem. Eng. Sci.}}
\textbf{\bibinfo{volume}{60}}, \bibinfo{pages}{57--72} (\bibinfo{year}{2005}).

\bibitem {ENSMeanderGranular2011}\bibinfo{author}{Varas, G.},
\bibinfo{author}{Vidal, V.} \& \bibinfo{author}{G{\'e}minard, J.-C.}
\newblock \bibinfo{title}{Venting dynamics of an immersed granular layer}.
\newblock \emph{\bibinfo{journal}{Phys. Rev. E}} \textbf{\bibinfo{volume}{83}}%
, \bibinfo{pages}{011302} (\bibinfo{year}{2011}).

\bibitem {FluidizedJet2015}\bibinfo{author}{Lungu, M.},
\bibinfo{author}{Wang, J.} \& \bibinfo{author}{Yang, Y.}
\newblock \bibinfo{title}{Numerical simulations of flow structure and heat
transfer in a central jet bubbling fluidized bed}.
\newblock \emph{\bibinfo{journal}{Powder Tech.}}
\textbf{\bibinfo{volume}{269}}, \bibinfo{pages}{139--152} (\bibinfo{year}{2015}).

\bibitem {BubbleEffectsFluidizedBed1981}\bibinfo{author}{Sit, S.} \&
\bibinfo{author}{Grace, J.}
\newblock \bibinfo{title}{Effect of bubble interaction on interphase mass
transfer in gas fluidized beds}.
\newblock \emph{\bibinfo{journal}{Chem. Eng. Sci.}}
\textbf{\bibinfo{volume}{36}}, \bibinfo{pages}{327--335} (\bibinfo{year}{1981}).

\bibitem {ShieldsNumPascale2007}\bibinfo{author}{Ouriemi, M.},
\bibinfo{author}{Aussillous, P.}, \bibinfo{author}{Medale, M.},
\bibinfo{author}{Peysson, Y.} \& \bibinfo{author}{Guazzelli, E.}
\newblock \bibinfo{title}{Determination of the critical shields number for
particle erosion in laminar flow}.
\newblock \emph{\bibinfo{journal}{Physics of Fluids}}
\textbf{\bibinfo{volume}{19}} (\bibinfo{year}{2007}).

\bibitem {midi2004dense}\bibinfo{author}{MiDi, G.}
\newblock \bibinfo{title}{On dense granular flows}.
\newblock \emph{\bibinfo{journal}{Eur. Phys. J. E}}
\textbf{\bibinfo{volume}{14}}, \bibinfo{pages}{341--365} (\bibinfo{year}{2004}).

\bibitem {cassar2005submarine}\bibinfo{author}{Cassar, C.},
\bibinfo{author}{Nicolas, M.} \& \bibinfo{author}{Pouliquen, O.}
\newblock \bibinfo{title}{Submarine granular flows down inclined planes}.
\newblock \emph{\bibinfo{journal}{Phys. Fluids}} \textbf{\bibinfo{volume}{17}}%
, \bibinfo{pages}{103301} (\bibinfo{year}{2005}).

\bibitem {du2003granular}\bibinfo{author}{du~Pont, S.~C.},
\bibinfo{author}{Gondret, P.}, \bibinfo{author}{Perrin, B.} \&
\bibinfo{author}{Rabaud, M.}
\newblock \bibinfo{title}{Granular avalanches in fluids}.
\newblock \emph{\bibinfo{journal}{Phys. Rev. Lett.}}
\textbf{\bibinfo{volume}{90}}, \bibinfo{pages}{044301} (\bibinfo{year}{2003}).

\bibitem {PouliquenBook}\bibinfo{author}{Andreotti, B.},
\bibinfo{author}{Forterre, Y.} \& \bibinfo{author}{Pouliquen, O.}
\newblock \emph{\bibinfo{title}{Granular media: between fluid and solid}}
(\bibinfo{publisher}{Cambridge University Press}, \bibinfo{year}{2013}).

\bibitem {HerminghausBook}\bibinfo{author}{Herminghaus, S.}
\newblock \emph{\bibinfo{title}{Wet Granular Matter: A Truly Complex Fluid}}
(\bibinfo{publisher}{World Scientific, Singapore}, \bibinfo{year}{2013}).

\bibitem {LiuNagel}\bibinfo{author}{Liu, A.~J.} \&
\bibinfo{author}{Nagel, S.~R.}
\newblock \bibinfo{title}{Nonlinear dynamics: Jamming is not just cool any
more}. \newblock \emph{\bibinfo{journal}{Nature}}
\textbf{\bibinfo{volume}{396}}, \bibinfo{pages}{21--22} (\bibinfo{year}{1998}).

\bibitem {TakeharaPRL2014}\bibinfo{author}{Takehara, Y.} \&
\bibinfo{author}{Okumura, K.}
\newblock \bibinfo{title}{High-velocity drag friction in granular media near
the jamming point}. \newblock \emph{\bibinfo{journal}{Phys. Rev. Lett.}}
\textbf{\bibinfo{volume}{112}}, \bibinfo{pages}{148001} (\bibinfo{year}{2014}).

\bibitem {BehringerNature94Fluidized}\bibinfo{author}{Pak, H.} \&
\bibinfo{author}{Behringer, P.}
\newblock \bibinfo{title}{Bubbling in vertically vibrated granular materials}.
\newblock \emph{\bibinfo{journal}{Nature}} \textbf{\bibinfo{volume}{371}},
\bibinfo{pages}{231--233} (\bibinfo{year}{1994}).

\bibitem {howard1989fluidized}\bibinfo{author}{Howard, J.}
\newblock \emph{\bibinfo{title}{Fluidized bed technology: principles and
applications}} (\bibinfo{publisher}{A. Hilger}, \bibinfo{year}{1989}).
\end{thebibliography}

\section*{Acknowledgements}

K. O. thanks Naoyuki Sakumichi (Ochanomizu University) for useful comments.
This research was partly supported by Grant-in-Aid for Scientific Research (A)
(No. 24244066) of JSPS, Japan.

\section*{Author contributions statement}

K. O. and Y. Yagisawa conceived the experiment, and Y. Yoshimura conducted the
experiments. Y. Yoshimura and K.O. analyzed the results and prepared the
figures and graphs, and K.O. wrote the manuscript. All authors reviewed the manuscript.

\section*{Additional information}

Competing financial interests: The authors declare no competing financial interests.

\clearpage

\end{document}